# Performance Analysis of 60nm gate length III-V InGaAs HEMTs: Simulations vs. experiments

Neophytos Neophytou[1], Titash Rakshit[2] and Mark S. Lundstrom[3]

[1,3]Network for Computational Nanotechnology, Purdue University, W. Lafayette, IN

[2]Intel Corporation

## Abstract

An analysis of recent experimental data for high-performance $In_{0.7}Ga_{0.3}As$ high electron mobility transistors (HEMTs) is presented. Using a fully quantum mechanical, ballistic model, we simulate $In_{0.7}Ga_{0.3}As$ HEMTs with gate lengths of $L_G$ = 60nm, 85, and 135 nm and compare the result to the measured I-V characteristics including drain-induced barrier lowering, sub-threshold swing, and threshold voltage variation with gate insulator thickness, as well as on-current performance. To first order, devices with three different oxide thicknesses and channel lengths can all be described by our ballistic model with appropriate values of parasitic series resistance. For high gate voltages, however, the ballistic simulations consistently overestimate the measured on-current, and they do not show the experimentally observed decrease in on-current with increasing gate length. With no parasitic series resistance at all, the simulated on-current of the $L_G$ = 60 nm device is about twice the measured current. According to the simulation, the estimated ballistic carrier injection velocity for this device is about $2.7 \times 10^7$ cm/s. Because of the importance of the semiconductor capacitance, the simulated gate capacitance is about 2.5 times less than the insulator capacitance. Possible causes of the transconductance degradation observed under high gate voltages in these devices are also explored. In addition to a possible gate-voltage dependent scattering mechanism, the limited ability of the source to supply carriers to the channel, and the effect of non-parabolicity are likely to play a role. The drop in on-current with increasing gate length is an indication that the devices operate below the ballistic limit.

# I. Introduction

Field-effect transistors with III-V channel materials have recently received much attention because of their potential as switching devices for future digital technology nodes. Both heterostructure based high electron mobility transistors (HEMTs) [1, 2, 3] and MOSFETs [4, 5] have been reported. Due to their higher mobility, the III-V channel materials should reach the ballistic limit at longer channel lengths than Si devices. The low effective mass of the III-Vs should also boost the ballistic carrier velocity and improve the $I_D$-$V_D$ characteristics. Trade-offs are involved (e.g. the light effective mass leads to a density-of-states bottle neck [6, 7] and to source-drain tunneling [8]), but III-V FETs have the potential to outperform Si MOSFETs under low-voltage operation. In that regard, high-performance HEMTs based on III-V compounds with channel lengths below 90 nm have recently been demonstrated [2, 3, 9, 10]. Good control of the wide bandgap insulator thickness down to 3nm was achieved while still maintaining relatively low gate leakage currents – even under high biases. This paper is a simulation study of the results reported by Kim *et al*. [2]. Our objective is to examine the experimental data with a fully quantum mechanical, ballistic model in order to understand what can and cannot be explained.

In this paper, a two-dimensional, ballistic quantum transport HEMT simulator based on the real space Non-Equilibrium Green's Function (NEGF) approach [11, 12] is employed. Simulation results show that for these $In_{0.7}Ga_{0.3}As$ HEMTs with a gate length of 60 nm and zero series resistance, a ballistic device of this kind would deliver about twice the on-current of the measured device. With external series resistors added, the simulated I-V characteristics are close to the measured results, except at the highest gate voltage. While the discrepancy at high gate voltages might be due to scattering, source design and conduction band non-parabolicity are equally likely explanations. The ballistic simulations show good agreement with the subthreshold swing and drain-induced barrier lowering vs. channel length, but they do not show the drop in on-current with increasing gate length that is observed experimentally. This suggests that scattering is important in the longer channel length devices. The simulations shed light on the internal physics of these devices and identify issues for further study.

## II. Approach

For simulation purposes, the device geometry was simplified as shown in Fig. 1. In the experimental device [2], the source and drain contacts are located on the top of the device, and the current flow is two-dimensional through a doped heterostructure stack. Rather than attempting to simulate the contacts (and the associated metal-semiconductor contact resistance), we placed ideal contacts at the two ends of the channel as shown in Fig. 1 and added extrinsic series resistors to the source and drain. The simulated HEMT consists of a 15nm $In_{0.7}Ga_{0.3}As$ layer between two $In_{0.52}Al_{0.48}As$ buffer layers. The gate electrode in the simulated device is placed on top of the $In_{0.52}Al_{0.48}As$ layer, which (in the simulated structure) has the same thickness throughout the entire length of the device. A silicon δ-doped layer in the $In_{0.52}Al_{0.48}As$ buffer layer effectively dopes the source/drain regions of the device to $2.1 \times 10^{12}/cm^2$ [13]. The δ-doped layer is located 3nm away from the channel layer. Devices with insulator thickness of $t_{ins}$=3nm, 7nm and 11nm were described in [2]. Later, better estimates of the 7nm and 11nm devices were given as 6.5nm and 10nm [13], and these were the values used in our simulations. When simulating the $t_{ins}$=3nm device, the δ-doped layer was placed on top of the insulator, and the gate electrode on top of a thin layer on top the δ-doped layer. In the simulation, the δ-doped layer is given a finite thickness of 0.40 nm and the thin layer on top of the d-doped layer was 0.40 nm. The result was an insulator thickness of $t_{ins}$=3.8nm in the simulation. This is within the experimental uncertainty in the insulator thicknesses of +/- 1nm [13]. The uncertainty in insulator thickness is not substantial for the thinnest insulator device and this is discussed in Sec. IV.

In the simulations, the far left/right regions of the δ-doped layer are doped to $1 \times 10^{13}/cm^2$ to mimic additional doping from the n+ cap layers ($In_{0.7}Ga_{0.3}As/In_{0.52}Al_{0.48}As$) used in the experimental device to facilitate ohmic contacts to the source and drain. There are, therefore, two different doping regions in the simulated device. The region that is directly adjacent to the channel to its left/right ($L_{side}$), has a carrier density of $2.1 \times 10^{12}/cm^2$, which is the value specified by the experimental group [13]. The far left/right region has a larger doping of $10^{13}/cm^2$. This level of doping is unrealistically high for this type of material, but it favors numerical stability of the simulation and does not affect the device. This is a way to mimic the extended source/drain regions of the actual device. In addition, although the lightly doped region in the experiment has $L_{side}$ = 1μm [13], for computational efficiency $L_{side}$ is set to $L_{side}$= 60nm in the simulations. As will be discussed later, under high gate bias, the source design becomes important. Because of the simplified source design used in the simulation, we will be able to draw only qualitative conclusions about the high gate bias performance. The channel region is the region directly under the gate electrode and has $L_G$ = 60 nm as in the experimental device. Longer channel lengths were also examined experimentally, and these devices are briefly considered in Sec. IV.

It should be pointed out that the source and drain contacts in the simulated device should not be regarded as real contacts with an associated contact resistance. Rather, they are idealized contacts to the extended source/drain regions, which are assumed to be maintained in thermodynamic equilibrium by strong scattering. Venugopal et al. examined this assumption for silicon transistors and found that scattering in typical contacts of heavily doped silicon is sufficient to maintain thermodynamic equilibrium [14]. Nevertheless, this assumption may need to be reconsidered as devices continue to shrink and for new channel materials such as the III-V's considered here. Indeed, Fischetti has discussed the phenomenon of "source starvation" which is a manifestation of non-equilibrium contacts in III-V FETs [15]. For this study, we assume extended contacts that are maintained in thermodynamic equilibrium.

Figure 1b shows the simulated conduction band profile normal to the channel taken at a location near the source end of the channel (near the top of the potential energy barrier between the source and the channel) when the device is under large gate bias. The workfunction difference between the gate and the $In_{0.52}Al_{0.48}As$ buffer layer is adjusted to $\Delta\Phi_B=0.5eV$ in order to match the threshold voltage of the simulated devices to the experimental measurements. The thickness of the $In_{0.52}Al_{0.48}As$ layer in this case is 3nm, and the δ-doped layer is adjacent to the gate/$In_{0.7}Ga_{0.3}As$ interface. The conduction band discontinuity between the In$_{0.7}$Ga$_{0.3}$As/In$_{0.52}$Al$_{0.48}$As layer is assumed to be $\Delta E_c$ = 0.6eV [13, 16]. The dielectric constant of In$_{0.52}$Al$_{0.48}$As is assumed to be $\varepsilon$ = 14 and that of the In$_{0.7}$Ga$_{0.3}$As is $\varepsilon$ = 14.5 [17].

The effective mass of the In$_{0.7}$Ga$_{0.3}$As channel is an input to the simulation. Because of conduction band non-parabolicity, quantum confinement will increase the effective mass as compared to its value in the bulk. In principle, the appropriate effective mass could be extracted from atomistic calculations (e.g. tight-binding atomistic methods), however, this is a difficult task because the masses can be a function of the exact placement of the atoms in the structure and the distortions within the structure. In this work a simplified approach is followed; we extract the effective mass from atomistic, tight-binding [18] calculations for a 15 nm wide In$_{0.7}$Ga$_{0.3}$As quantum well structure without assuming any lattice distortions. The dispersion of the quantum well is shown in Fig. 2. The wafer orientation is (100) and the transport orientation is [011]. The parabolic band drawn on top of the first valley is adjusted to match the density of states up to 0.2eV above the conduction band edge and results in an effective mass of $m^*$ = 0.048m$_0$, which is the value used in the simulations. (We chose to fit from the bottom of the conduction band to 0.2 eV above the bottom because the maximum position of the Fermi level above the conduction band edge under high gate bias is usually close to or below 0.2eV). Similar parabolic bands that match the *bulk E(k)* bandstructures of the InAs and GaAs at $\Gamma$ up to 0.2eV above the conduction band minima were also extracted. A weighted average of these masses according to the 70% indium and 30% gallium composition results in a very similar value for the effective mass. The mass value is higher than the weighted average of the literature bulk masses, which is $m^*$ = 0.037m$_0$, ($m^*_{InAs}$ = 0.027m$_0$, $m^*_{GaAs}$ = 0.063m$_0$). Our use of a larger effective mass accounts for the effect of non-parabolicity in an approximate way. In addition, as shown in Fig. 2, the L valleys are very high in energy compared to the $\Gamma$ valleys and are, therefore, ignored in our simulations. (This is expected since the composite channel in this case due to the 70% indium composition has stronger InAs properties rather than GaAs properties which will tend to place the $\Gamma$ and L valleys closer in energy.)

The non-equilibrium Green's function approach [11, 12] for ballistic quantum transport is self consistently coupled to a 2D Poisson solver for treatment of the electrostatics. Since the channel is relatively thick (15nm), significant potential variations are expected in the cross section along the transport orientations. The NEGF Hamiltonian uses a real space technique in the parabolic, effective mass approximation (EMA) and accurately accounts for the mode coupling when large potential variations exist. The NEGF transport equation is solved in the channel area as well as in the upper $In_{0.52}Al_{0.48}As$ buffer layer in order to capture the wavefunction penetration in that layer. The 2D Poisson's equation is solved in the entire cross section of the device in order to accurately capture the 2D electrostatics of the device.

### III. RESULTS

In order to compare the measured to the simulated data, two fitting parameters were used, a value of external series resistance ($R_{SD}$) that is added to the device, and the workfunction difference ($\Delta\Phi_B$) between the gate and the $In_{0.52}Al_{0.48}As$ buffer layer. The effect of the series resistance will be explained in the following section. The workfunction difference, $\Delta\Phi_B$, is used to adjust the $V_T$ of the simulated to that of the experimental data. The adjustment is done once for the $L_G$ = 60 nm device with a 3nm thick insulator. The result, $\Delta\Phi_B = 0.5eV$, is a reasonable number for the workfunction difference between the two materials. Figure 3a shows the $I_D$-$V_G$ characteristics for the three devices – each with a 60nm channel length but with three different insulator thicknesses. The measured and simulated curves agree fairly well. In the case of the 7nm insulator device, the simulated and measured $V_T$ differs by ~0.04V. This small deviation might be due to various reasons such as interface traps, charged impurities, or uncertainties in the thickness of the layers in the experimental device. As the insulator thickness increases, there is a large negative shift in the $V_T$ by almost 0.25V, which is attributed to the δ-doped layer and its increasing effect on the electrostatics of the channel as the gate electrode moves farther away. The threshold voltage shift is well-described by

$$\Delta V_T = \frac{qN_W}{C_{ins}} \frac{\hat{x}}{T_{ins}}, \quad (1)$$

where $N_W$ is the δ-doping concentration per cm$^2$, and $\hat{x}$ is the centroid of the charge distribution in the insulator [19].

The Drain Induced Barrier Lowering (*DIBL)* and the subthreshold swing (SS) extracted from the simulated data are shown in Fig. 3b, and both are seen to increase as the insulator thickness increases, which is expected from 2D electrostatics. The simulated results agree with the experimental data both qualitatively and quantitatively.

The second adjustable parameter in the simulation is the series resistance. The series resistance originates from the complicated ohmic contact between the n$^+$ cap layer, the In$_{0.52}$Al$_{0.48}$As layer and the barrier between the interface of the In$_{0.52}$Al$_{0.48}$As/ In$_{0.7}$Ga$_{0.3}$As layers. Figure 4a shows the experimental $I_D$-$V_D$ data and the simulated ballistic $I_D$-$V_D$ characteristic at the same gate overdrive ($V_G$ = 0.5V). The simulated, ballistic ON-current is almost double than the experimental value, and the channel resistance of the simulated ballistic device is $R_B$ = 170 Ω-μm (inverse slope of the linear region). In order fit the simulated results to the experimental data, a series resistance (source plus drain) of $R_{SD}$ = 400 Ω-μm was added to the ballistic data in order to match the total resistance measured in the experimental data (inverse slope of the high $V_G$ experimental $I_D$-$V_D$).

Once the series resistance is fit to the linear region of the highest gate voltage data, the simulated data at low drain voltages shows very good agreement with the experimental observations for all three gate bias cases reported experimentally ($V_G$ = 0.1V, 0.3V, 0.5V). The agreement at high drain voltage is also good, except for a ~15% discrepancy between ON-current of the measured and simulated data. For this $L_G$ = 60 nm device, the experimental results can, to a reasonable approximation, be explained by an intrinsic, ballistic FET with two series resistors attached to it, except for the overestimate of the on-current, which will be discussed in Sec. IV. Longer channel lengths appear to operate at a lower fraction of the ballistic limit, as will also be discussed in Sec. IV.

Similarly, the experimental $I_D$-$V_D$ data for the $t_{ins}$ = 7 nm and $t_{ins}$ = 11nm devices can be explained by using slightly different values of $R_{SD}$ ($R_{SD}$ = 350 Ω-μm and $R_{SD}$ = 310 Ω-μm, respectively). The value of the fitted series resistance increases as the insulator thickness decreases. This was also observed in the experiments and was attributed to the isotropic etching that was used before gate deposition to produce the three different insulator thicknesses [2]. As more of the insulator sidewall is etched, the series resistance tends to increase. Figures 4b,c show the experimental $I_D$-$V_D$ for various $V_G$ values compared to the simulated results after the series resistance has been fit. Good agreement between the experimental and simulated data is observed, but for each of the three cases, the ON-current of the simulated device is ~10-15% more than that of the measured device.

The mobility of a field-effect transistor is often extracted from the linear region current. Although mobility has no physical meaning in our ballistic simulations, the simulated ballistic drain current is linearly proportional to the drain voltage at low $V_{DS}$, so we can extract a "mobility" by equating the channel resistance to a conventional MOSFET expression,

$$R_{ch} = \frac{V_{DS}}{I_{DS}/W} \equiv \frac{L}{\mu_B C_{ins}(V_G - V_T)}, \tag{2}$$

where $\mu_B$ is the so-called ballistic mobility by [20, 21, 22]. From our simulations, $R_{ch}$ at high gate bias (before adding the effect of $R_{SD}$), varies between $R_{ch}$=170 Ω-μm – 240 Ω-μm as the insulator thickness varies from 3nm to 11nm. From these channel resistances, a value of the ballistic mobility is extracted to be $\mu_B$ ~ 170-450 cm$^2$/V-s. Although the mobility of bulk In$_{0.7}$Ga$_{0.3}$As is measured to be ~10,000 cm$^2$/V-s, the "apparent" mobility (in the sense of eqn. (2)) that a short channel HEMT can display is limited to a few hundred. Alternatively, one could deduce a mobility for the device by plotting the total resistance between the source and drain as a function of channel length. The y-intercept of this curve would be the fixed, external series resistance and the inverse of the slope would be proportional to the channel mobility. In that case, a ballistic FET would show zero slope, corresponding to an infinite mobility.

## IV. DISCUSSION

Within the uncertainties of the simplified structure used in the simulations and in our knowledge of various device parameters, the results presented in the previous section show that the $L_G$ = 60 nm HEMTs reported by Kim et al. [2] can be approximately described as ballistic FETs with two external series resistors. The only significant discrepancy between the simulated and experimental results is the consistent 10-15% over-estimate of the ON-currents. The experimental transconductance, $g_m$, vs. gate voltage characteristic is shown in Fig. 5 for the $t_{ins}$= 3nm device. The observed degradation in $g_m$ at high gate voltages might be attributed to various causes. Scattering at high gate biases could reduce mobility and degrade $g_m$. Another possibility is population of heavy effective mass upper valleys. Figure 2 shows, however, that the L valleys are too high in energy to be populated. Parallel conduction in the upper layer, which has much heavier masses (~5 times heavier) than the channel layer, could also be a possibility. As shown in Fig. 1b, however, our simulations show no significant wavefunction penetration in the upper layer – even under high inversion conditions. Series resistance could be yet another possibility. Figure 5 shows the simulated $g_m$ vs. $V_G$ characteristics for three different values of series resistance ($R_{SD}$ = 0, 400, 800 Ω-μm). For the $R_{SD}$ = 0 and 400 Ω-μm cases, the $g_m$ follows the experimental curve, but saturates at much higher $V_G$ than the experimental curve. For the 800 Ω-μm characteristic, we obtain roughly the correct magnitude of $g_m$, but this value of $R_{SD}$ is too large to be consistent with the experimental measurement. The fact that $g_m$ degradation occurs even in the ballistic simulation tells us, however, that there might be other possibilities. Two other plausible causes, the design of the source, and the effects of non-parabolicity are discussed below.

For III-V transistors, the design of the source can be an important factor [15, 23]. Transistors operate by modulating potential energy barriers [24, 25]. As the gate voltage increases, the potential energy barrier decreases, and the charge in the channel increases. When the gate voltage increases to the point where the barrier is removed and the channel charge is equal to the charge in the source, transistor action degrades significantly.

Simply stated, there can't be more charge in the channel than in the source. For the HEMT under consideration here, the charge in the source ($2.1 \times 10^{12}/cm^2$), is much lower than typical for Si MOSFETs, so these source exhaustion effects become apparent at relatively low gate voltages.

Source design limits are illustrated by the ballistic simulation shown in Fig. 6. Figures 6a, b, c show the energy-resolved current vs. position for the HEMT device under different gate voltages. The conduction and valence bands are indicated (white-dot lines), and the current flows above the top of the conduction band. The source/drain regions consist of two portions, an $n^{++}$ region near the ideal contacts and an $n^+$ region adjacent to the channel. Figure 6a shows the OFF-state of the device, where the source Fermi level ($E_{fs}$) is well below the top of the source to channel energy barrier. As $V_G$ increases, the barrier in the channel decreases – eventually reaching the same level as the n+ source region (Fig. 6b). The top of the barrier has in this case shifted to the beginning of the $n^{++}$ source region. When $V_G$ increases even more (Fig. 6c), the gate can only modulate the energy barrier at the $n^{++}$ to $n^+$ junction through weak fringing fields. Transistor action is lost, and $g_m$ drops as shown in Fig. 5 for both the simulated and measured characteristics. In our simulations, these effects are exaggerated by the assumption of ballistic transport in the $n^+$ source, but the effect is primarily an electrostatic one and is also observed in drift-diffusion simulations [26].

The gate voltage at which the transconductance begins to degrade is strongly dependent on the barrier between the channel and the source, which depends on the doping of the source. Figure 7 shows the simulated $g_m$ for structures with different δ-doping densities above the source/drain. As the doping in the source decreases, this effect shows up at smaller gate voltages. The low gate bias part of the $g_m$ vs. $V_G$ characteristic is not doping dependent because under low gate voltage, the source is able to supply the charge demanded by the gate voltage. In the experimental results, the $n^+$ source region was $L_{side}$ = 1μm in length, whereas in our simulation, $L_{side}$ = 60nm was used. The differences in the source doping profiles may explain why the transconductance is experimentally observed to degrade ~0.2V before the simulated transconductance.

Although we cannot unambiguously conclude that the observed transconductance degradation is due to source exhaustion, our simulations do clearly demonstrate that source design is an important issue for III-V MOSFETs. Finally, note that the effects discussed here are purely electrostatic in nature and occur in both ballistic and drift-diffusion simulations. Fischetti has discussed "source starvation," which results from a difficulty in injecting carriers into longitudinal momentum states in the channel [15]. Those effects were not included in our study and would only make source design an even more important issue.

Two important parameters for a FET are the charge and velocity at the beginning of the channel. Two questions arise. The first is: How close is the charge at the top of the potential barrier to the equilibrium MOS capacitor value of $Q = C_G(V_G - V_T)$? The second question is: How the velocity extracted from the numerical simulator compares to the ballistic injection velocity expected from the bandstructure of the channel. To answer both of these questions, the top of the potential barrier in the numerical results needs to be identified. Doing so is not as trivial, because of the large variation of the $E_C$ across the depth of the 15nm channel width. We employ two different methods to locate the top of the barrier. The first is to take the weighted average of the charge distribution with the 2D $E_C(x,y)$ profile with the 2D charge density $n(x,y)$ according to

$$\langle E_C(x) \rangle = \frac{\int n(x,y) E_C(x,y) dy}{\int n(x,y) dy}. \tag{3}$$

Figure 8a shows the resulting $\langle E_C(x) \rangle$ (white-dotted line) superimposed on the electron density spectrum plot. Figure 8a is plotted at $V_G = 0.4$V, and $V_D = 0.35$V, which are the estimated intrinsic device voltages at the ON-state (after accounting for the effect of $R_{SD}$). From Fig. 8a, the top-of-the-barrier can be identified to reside at 105nm (5nm inside the channel from the point where the gate electrode begins).

A second way to identify the top-of-the-barrier is by identifying the point of maximum gate control by locating the position where $dN_S(x)/dV_G$ is maximized (where $N_S(x)$ is the charge in the channel per cm$^2$). This method places the top-of-the-barrier at 104 nm. Both approaches give very similar results, so we take the top-of-the-barrier to be at 104.5nm. The corresponding charge and the velocity (defined as $I_{ON}/N_S(x)$) at the-top-of-the-barrier are $N_S \approx 1.3 \times 10^{12}$ per cm$^2$ and $\upsilon_{ave} \approx 2.7 \times 10^7$ cm/s as shown in Fig. 8b,c respectively. The charge density and velocity are rather low for this light mass channel due to the fact that the source Fermi level is less than 0.1eV above $E_C$ under ON-state conditions. Figure 8 shows that these quantities are very sensitive to the precise location of the beginning of the channel. This information is available in our simulator, but it is not available when analyzing experimental data.

To answer the first question about how close the charge is to $Q = C_G(V_G - V_T)$, the simulated equilibrium carrier density vs. gate voltage is plotted in Fig. 8d (solid-blue). The quantity $Q = C_{ins}(V_G - V_T)$ with $C_{ins}$= 0.032 F/m$^2$ (or $3.2 \times 10^{-6}$ F/cm$^2$) is shown as the solid-square-black line of Fig. 8d. Assuming that $C_G = C_{ins}$ clearly over-estimates the charge. From the slope of the $C_G$ vs. $V_G$ plot (dashed-red line), we observe that $C_G$ is 2.5 times smaller than $C_{ins}$. From $C_G = C_{ins}C_S/(C_{ins} + C_S)$, we obtain a semiconductor capacitance of $C_S = 0.67 C_{ins}$. A simple calculation of the quantum capacitance, however, shows that $C_Q \sim 1.5 C_{ins}$, which indicates that $C_S$ is a factor of ~2 less than $C_Q$. As discussed by Pal [27], this occurs when the shape of the quantum well is bias-dependent.

According to Fig. 8d, at $V_{GS}$ = 0.4V, the charge at the top of the barrier under equilibrium conditions is $N_S \approx 1.5 \times 10^{12}$ per cm$^2$. The value found from the simulation under $V_{DS}$ = 0.35V is $N_S \approx 1.3 \times 10^{12}$ per cm$^2$, which is lower than the equilibrium value. It might be expected that *DIBL* would reduce $V_T$ and therefore increase the charge. Part of the reason for the lower charge under drain bias could be that only the positive velocity states are occupied at high $V_D$. The quantum capacitance, therefore, decreases under large drain bias by a factor of two. The lower $C_Q$ lowers the semiconductor capacitance $C_S$ and offsets the *DIBL*. The result is that the charge at the top of the barrier is somewhat less under high $V_{DS}$.

The second question had to do with the value of the ballistic velocity from the numerical simulation as compared to the value expected from the bandstructure. For a given $E(k)$ and Fermi level, we can determine the corresponding $N_S$ and $\langle v \rangle = v_{ave}$ under ON-state conditions where only $+k$ states are occupied. Figure 9 shows the result for the parabolic effective mass (EMA) dispersion used in the quantum simulations (square-blue). For comparison, the InAs and GaAs velocities are shown, calculated using dispersions extracted from an atomistic tight-binding model [18]. The weighted average of these two results is also shown in Fig. 9 (solid-brown). The weighted average tight-binding results resemble the effective mass results for the $In_{0.7}Ga_{0.3}As$ channel. The EMA velocity is in good agreement with the "weighted average" curve at low carrier densities, but at higher densities, the EMA velocity is higher, because non-parabolicity reduces the velocity in the tight-binding model. At an inversion charge density of $N_S = 1.3 \times 10^{12}$ per $cm^2$, which corresponds to the charge at the top of the barrier in the numerical simulation, the velocity for the EMA is $v_{ave} = v_{inj} \approx 4 \times 10^7$ cm/s, while for the weighted average tight-binding curve it is $v_{ave} = v_{inj} \approx 3.6 \times 10^7$ cm/s. These values are both higher than the $v_{ave} \approx 2.7 \times 10^7$ extracted from the NEGF simulation

The difference in the velocities deduced from the bandstructure and that extracted from the NEGF simulation might have to do with tunneling currents and quantum mechanical reflections around the top-of-the-barrier, which tend to reduce the average velocity. (In support of this conjecture, we note that the *Fermi level* in the quantum model is almost a $k_BT$ closer to $E_C$ than in the semiclassical model at the *same* carrier density, which indicated a carrier population below the top-of-the-barrier, and/or "negative" going state population in the quantum model). It is also evident in Fig. 9 that nonparabolicity can be important at this bias regime and can cause about 10% degradation in the average carrier velocity. Nonparabolity is another possible contribution to the $g_m$ degradation observed in the experimental data but not captured in the EMA treatment.

The main analysis of the discussion section up to now considered the $t_{ins}$ = 3 nm and $L_G$ = 60 nm device. The experimental data show variations in both changes in the InAlAs insulator thickness as well as gate length dependence. These two issues are briefly discussed here. Figure 10a shows how the insulator thickness affects the performance of the $L_G$ = 60 nm device. The equilibrium carrier density in the channel under $V_{GS}$ = 0.4V is shown in solid-square-blue, extracted as in Fig. 8d for all devices at the same $V_G$ - $V_T$. The carrier density in the channel doubles as the insulator thickness decreases from $t_{ins}$ = 10 nm to $t_{ins}$ = 3 nm – as expected. Under a high drain bias of $V_D$ = 0.35V, however, the carrier density at the top of the barrier (dash/dot-diamond-black) shows a much slower variation with insulator thickness. This occurs because under high drain biases $C_Q$ decreases by a factor of ~2, which drives the device toward the quantum capacitance limit in which variations in $C_{OX}$ are not as significant. Increasing DIBL with increasing insulator thickness lowers the $V_T$ and increases the charge in the channel. An interplay between these two effects reduces the charge variations as a function of $t_{ins}$. The increase in charge as the $t_{ins}$ is scaled from 10nm to 3nm is only ~30%. The velocity at the top of the barrier (dash-circle-red) shows an increase of ~20% with insulator thickness scaling. Scaling the insulator thickness down to 3nm can, therefore improve performance. Further scaling of the insulator, however, might not offer additional advantage at the on-state. Figure 10b shows the effect of scaling the insulator from $t_{ins}$=3.8nm to $t_{ins}$=3.0nm. This figure is the same as Fig. 4a, with the $t_{ins}$=3.0nm result also shown in black-diamond, plotted at the same $V_G$-$V_T$. The difference at the on-state is less than 5%.

We next investigate the gate length dependence of the $t_{ins}$ = 3 nm HEMTs. Experimentally in [2], $L_G$ = 60nm, 85nm and 135nm devices were reported. Significant gate length dependence was observed experimentally, with the ON-current decreasing as the gate length increases. This trend is shown in Fig. 11. This figure is the same as Fig. 4a, with all the three gate length data included (for clarity, we have shown only the highest gate voltage in each case). The solid-circle-red lines present the experimental data for the different gate lengths and for $V_G$ = 0.5V, 0.3V, and 0.1V. The solid-blue lines present the simulated results for the same devices after the series resistance was included. Although it is not shown in the plots, a good match was observed between the simulated and measured data for lower gate biases. For the high gate bias case, the simulated results show little gate length dependence – as it is expected from a ballistic model. The small differences originate from the changes in the electrostatics. The measured high $V_G$ data, however, show a significant gate length dependence. The longest device is about 40% below the ballistic simulation while the shortest device is only ~15% below. These results indicate increased scattering in the $L_G$ = 85nm device and even stronger scattering in the $L_G$ = 135nm device.

Finally, we should mention once again some of the uncertainties and simplifications that affect our analysis. The first is the ±1nm uncertainty in the etched AlInAs layer thickness, which however does not introduce considerable uncertainty at the ON-state. Second, the simplified device structure for the simulation had the source/drain regions that were only 60 nm long rather than 1 μm as in the experimental device. This simplification is likely to affect the high current region, where source design issues are expected to become important. Lattice distortions and the effect of strain in the channel were not considered and may have an impact on the effective mass of the channel.

## V. Conclusion

The performance of recently demonstrated high-performance $In_{0.7}Ga_{0.3}As$ HEMTs was investigated using a quantum ballistic model self consistently coupled to a 2D Poisson solver for electrostatics. With the addition of external series resistors, reasonable agreement between the ballistic simulation and the experimental data was obtained for all of the 60 nm channel length devices with insulator thicknesses of 3nm, 7nm, and 11nm using values of series resistance consistent with those measured in the experiments. Despite the simplifications in the model and the uncertainties in the exact values of the insulator thickness, series resistance and channel effective masses, these results suggest that 60nm channel length III-V HEMTs operate rather close to the ballistic limit. The on-current performance of longer channel lengths HEMTs, however, appears to be degraded by scattering although they still operate at over one-half of the ballistic limit.

For operation near the ballistic limit, the ballistic injection velocity rather than bulk mobility becomes the parameter of interest. The ballistic injection velocity for this device was found to be relatively low for this light effective mass material, because of the relatively low inversion charge operating conditions, quantum tunneling and reflections, and conduction band non-parabolicity. The semiconductor capacitance also plays an important role by increasing the effective oxide thickness (EOT) of the thinnest insulator device by 2.5 times. The results reported here strongly suggest that source design is an important factor for III-V FETs, as has also been recently pointed out by Fischetti [15]. These simulations also identify key factors for improving III-V HEMT performance as reduction of the parasitic series resistance, optimization of the source design, and reduction of the insulator thickness, which will be beneficial to the off-state performance but have only a small effect on the on-state performance.

## Acknowledgements

This work was funded by the Semiconductor Research Corporation (SRC) and by the Focus Center for Materials, Structures, and Devices. Computational resources for this work were provided through nanoHUB.org by the Network for Computational Nanotechnology (NCN). The authors would like to thank Dr. D.-H. Kim, Prof. J. del Alamo and Prof. Dimitri Antoniadis of the Massachusetts Institute of Technology for providing details about the experimental structures and for extensive discussions about the analysis presented here. Siyu Koswatta, Himadri Pal and Yang Liu are also acknowledged for helpful discussions, and Prof. Gerhard Klimeck is acknowledged for advice on the tight-binding calculations.

Figure captions

Figure 1:

(a) The simplified HEMT device structure. An $In_{0.7}Ga_{0.3}As$ between two $In_{0.52}Al_{0.48}As$ layers acts as the channel. A δ-doped layer 3nm away from the channel layer, effectively dopes the source/drain regions of the device to $2.1 \times 10^{12}/cm^2$ [13]. Heavier doping is used at the far left/right of the device. (b) The conduction band profile taken at a cross section of the HEMT device at the region of the source/channel boundary when the device is under large gate bias. The workfunction difference between the gate and the $In_{0.52}Al_{0.48}As$ buffer layer is adjusted to $\Delta\Phi_B=0.5eV$. The conduction band discontinuity between the $In_{0.7}Ga_{0.3}As/In_{0.52}Al_{0.48}As$ layer is assumed to be $\Delta E_c=0.6eV$. The dielectric constant of $In_{0.52}Al_{0.48}As$ is assumed to be $\varepsilon = 14$ and of the $In_{0.7}Ga_{0.3}As$ $\varepsilon=14.5$.

Figure 2:

The dispersion of the composite 15nm thick $In_{0.7}Ga_{0.3}As$ structure calculated using atomistic tight-binding calculations with no distortions taken into account. The wafer orientation is (100) and the transport orientation is [011]. The parabolic band (red-dotted) of $m^*=0.048$ $m_0$ is adjusted to match the density of states up to 0.2eV above the conduction band edge.

Figure 3

(a) The experimental (red-circle) and simulated (blue-solid) $I_D$-$V_G$ data for the $L_G=60nm$, $t_{ins}$=3nm, 7nm, 11nm devices. A workfunction difference between the gate and the $In_{0.52}Al_{0.48}As$ layer of $\Delta\Phi_B=0.5eV$ is used in order to match the $V_T$ for all devices. A negative shift in $V_T$ by 0.25V is observed as the oxide thickness increases. (b) The *DIBL* and subthreshold swing *(SS)* of the experimental and simulated data.

## Figure 4

Comparison between the experimental (red-circle) and simulated $I_D$-$V_D$ with series resistance added to them (blue-solid). (a) The $t_{ins}$=3nm device. Data for $V_G$ = 0.1V, 0.3V and 0.5V are shown. The black-dashed curve indicates the ballistic $I_D$-$V_D$ at $V_G$ = 0.5V with $R_{SD}$ = 0 Ω-μm. A $R_{SD}$ = 400 Ω-μm is added to the simulated data. (b) The $t_{ins}$=7nm device. Data for $V_G$ = 0V, 0.2V and 0.4V are shown. A $R_{SD}$ = 350 Ω-μm is added to the simulated data. (c) The $t_{ins}$=11nm device. Data for $V_G$ = -0.1V, 0.1V and 0.3V are shown. A $R_{SD}$ = 310 Ω-μm is added to the simulated data.

## Figure 5

The $g_m$ vs. $V_G$ data for the $t_{ins}$ = 3nm device. Measured data (red-circle), and simulated data with $R_{SD}$=0 Ω-μm (black-solid), $R_{SD}$ = 400 Ω-μm (blue-square), and $R_{SD}$ = 800 Ω-μm (green-square) are shown.

## Figure 6

The source exhaustion mechanism. The energy resolved current spectrum is shown. (a) The device at OFF-state. (b) The barrier collapses as $V_G$ is applied at ON-state. (c) Further increase in $V_G$ causes the lightly doped region to collapse. The top-of-the-barrier that has now shifted to the highly doped region and the gate loses control over the device.

## Figure 7

The effect of source/drain electron charge on $g_m$ degradation. As the "doping" decreases the degradation starts in lower gate biases.

## Figure 8

The intrinsic device parameters at ON-state. (a) The electron density spectrum. The density weighted $E_C$ and $E_V$ profiles are shown (dot-white lines). The top of the barrier is identified at 104.5nm. (b) The charge density along the length of the channel. (c) The average velocity along the length of the channel. (d) The equilibrium ($V_D$=0V) carrier density vs. $V_G$ (solid-blue). The charge as $C_{ins}*(V_G-V_T)$ is shown in solid-square-black. The charge as $(C_{ins}/2.5)*(V_G-V_T)$ is shown in dot-red.

## Figure 9

The "positive going" average bandstructure velocity vs. inversion carrier density of a 15nm thick quantum well, using a simple semiclassical ballistic model. The velocities of InAs and GaAs are shown in solid-square-black. Their bandstructures are calculated using an atomistic tight-binding model. The EMA bandstructure velocity for the dispersion used in the quantum simulation is shown in solid-circle-blue. The weighted average of the InAs and GaAs ($In_{0.7}Ga_{0.3}As$) velocity is shown in solid-brown.

## Figure 10:

(a) The simulated carrier density and average velocity at the same $V_G-V_T$=0.2V as a function of insulator thickness for the $L_G$=60nm device. Carrier densities for the $V_D$=0V (solid-square-blue) and $V_D$=0.35V (dash/dot-diamond-black) are presented. The gate bias is $V_G$=0.4V. The $V_G-V_T$=0.2V is the same for all insulator thickness devices at $V_D$=0V. No further $V_T$ adjustment was performed for the $V_D$=0.35V case. The average velocity (dash-circle-red) is calculated at $V_D$=0.35V. (b) The simulated and measured data are presented in a similar way to Fig. 4a for $V_G$=0.1V 0.3V and 0.5V and for $t_{ins}$=3nm (black-diamond), and 3.8nm (blue-solid). Variations in the insulator thickness do not introduce significant variations in the on-current.

## Figure 11:

Gate length dependence of the 3nm oxide thickness device. The simulated and measured data are presented in a similar way to Fig. 4a for $V_G$=0.1V 0.3V and 0.5V and for $L_G$=60nm, 85nm, 135nm. The simulated and measured data are in good agreement for the lower gate bias cases. Significant deviation is observed for the high bias cases, which is reduced as the gate length reduces. The simulated data do not show significant gate length dependence.

Figure 1: Device description

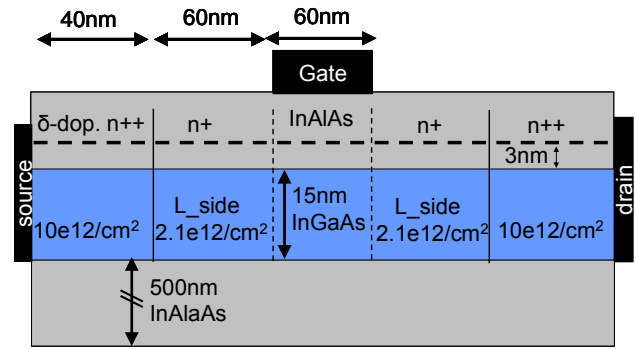

(a)

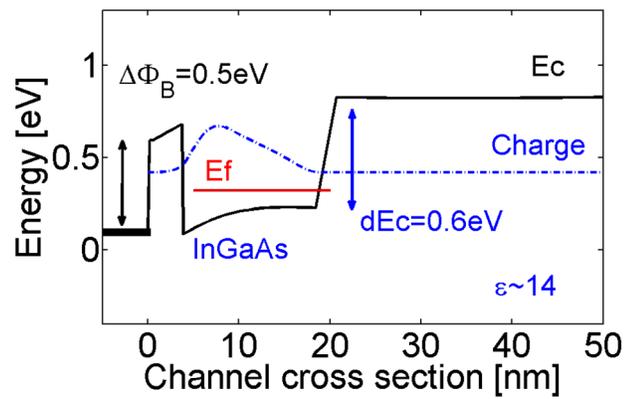

(b)

Figure 2: The E(k)

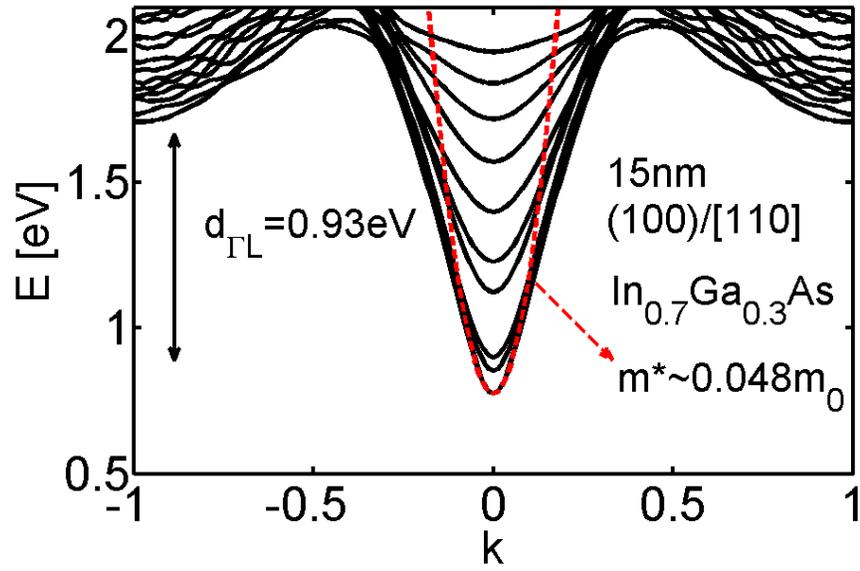

Figure 3: VT, DIBL, SS

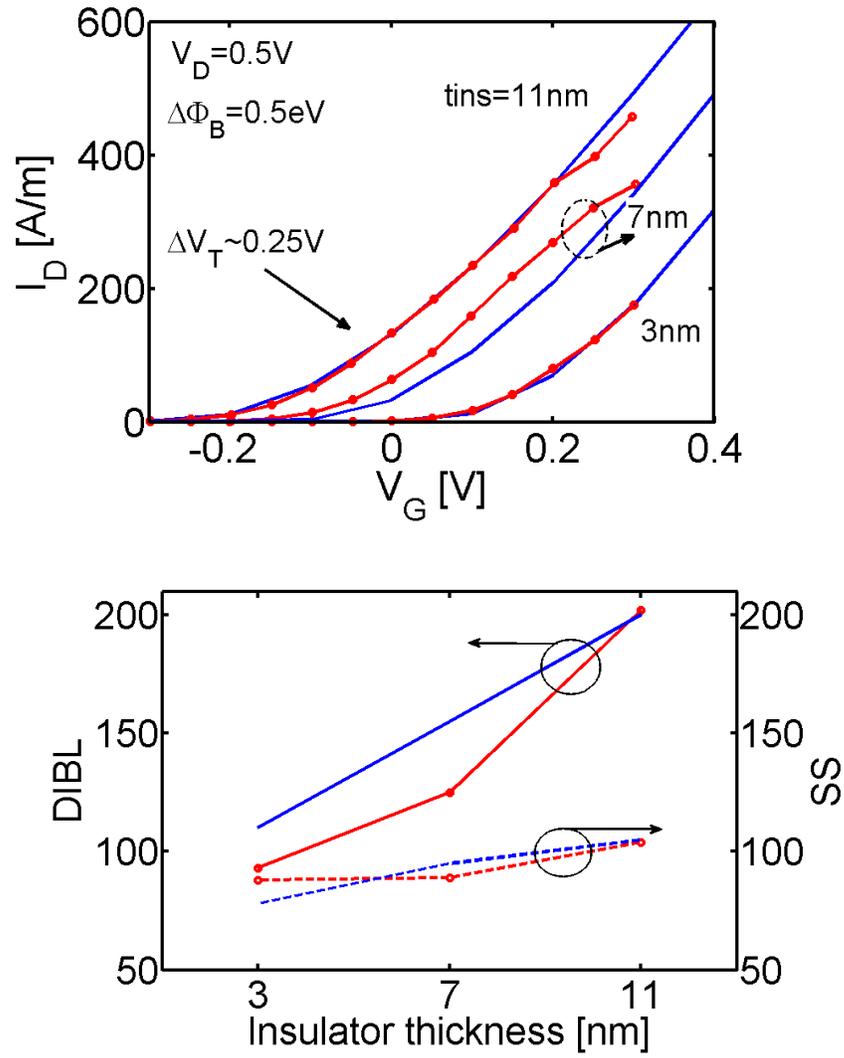

Figure 4: Series resistance - Ballistic mobility, ID-VD

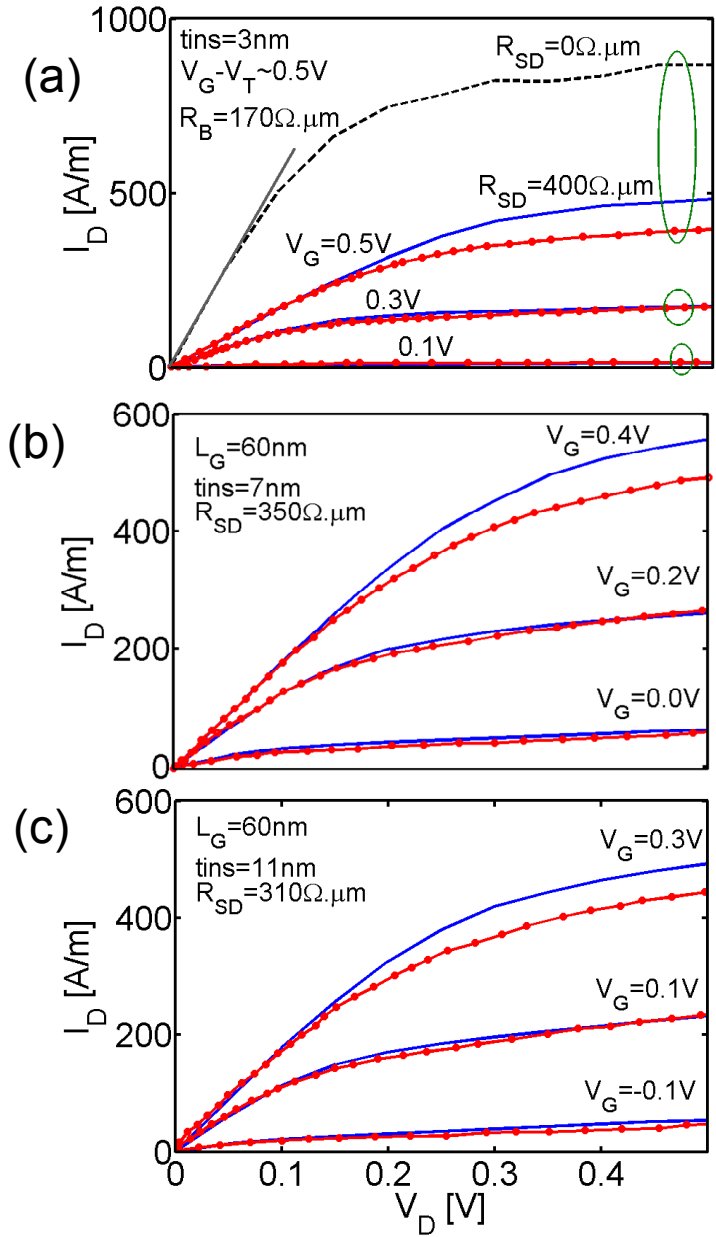

Figure 5: The effect of $R_{SD}$ on Gm

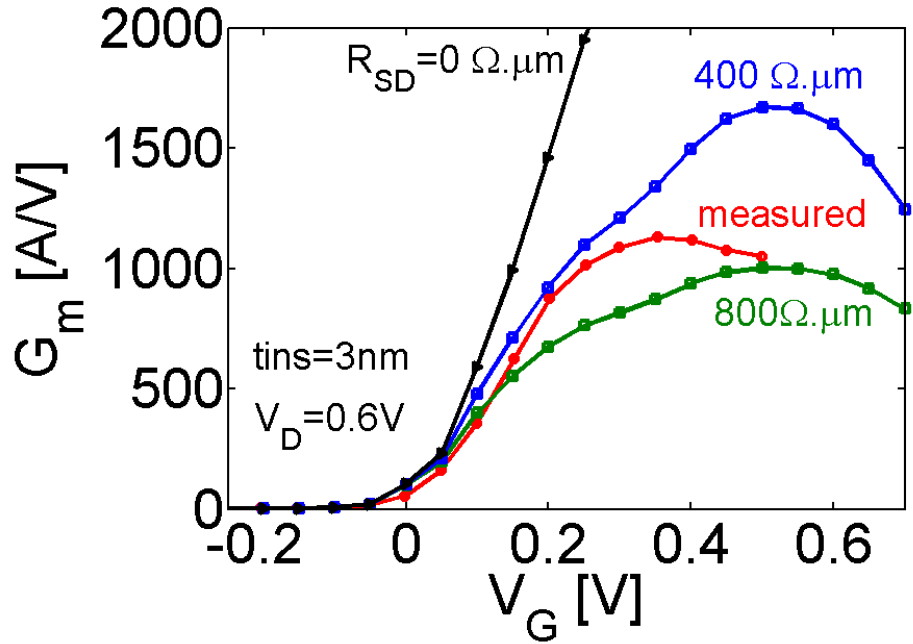

Figure 6: Source "exhaustion" as a reason for Gm degradation

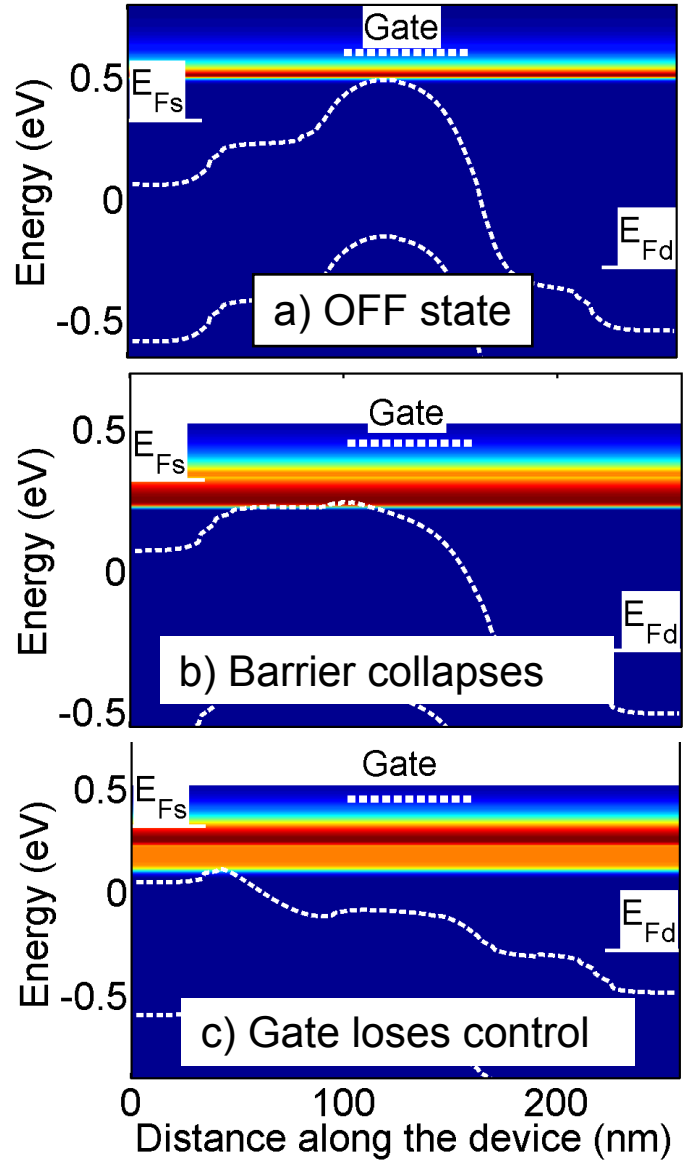

Figure 7:

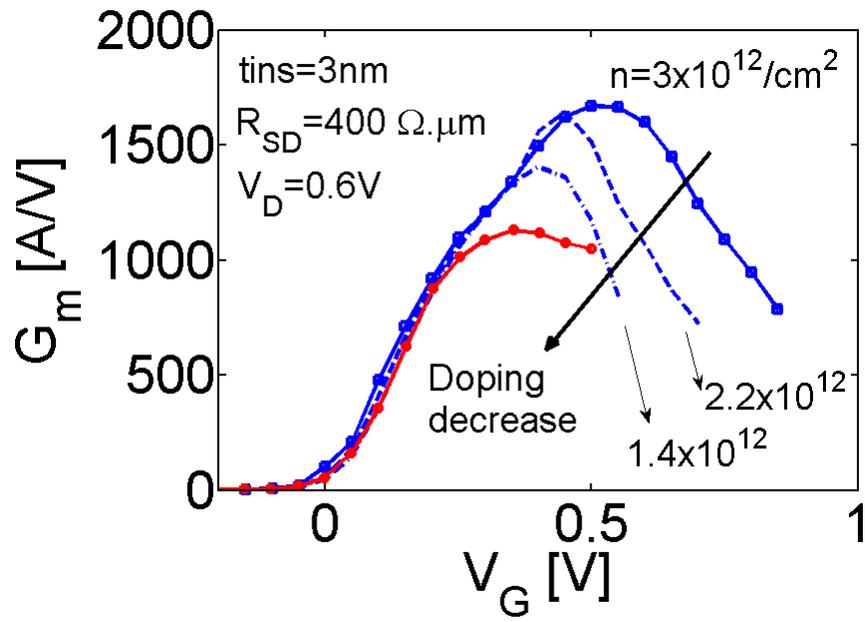

Figure 8

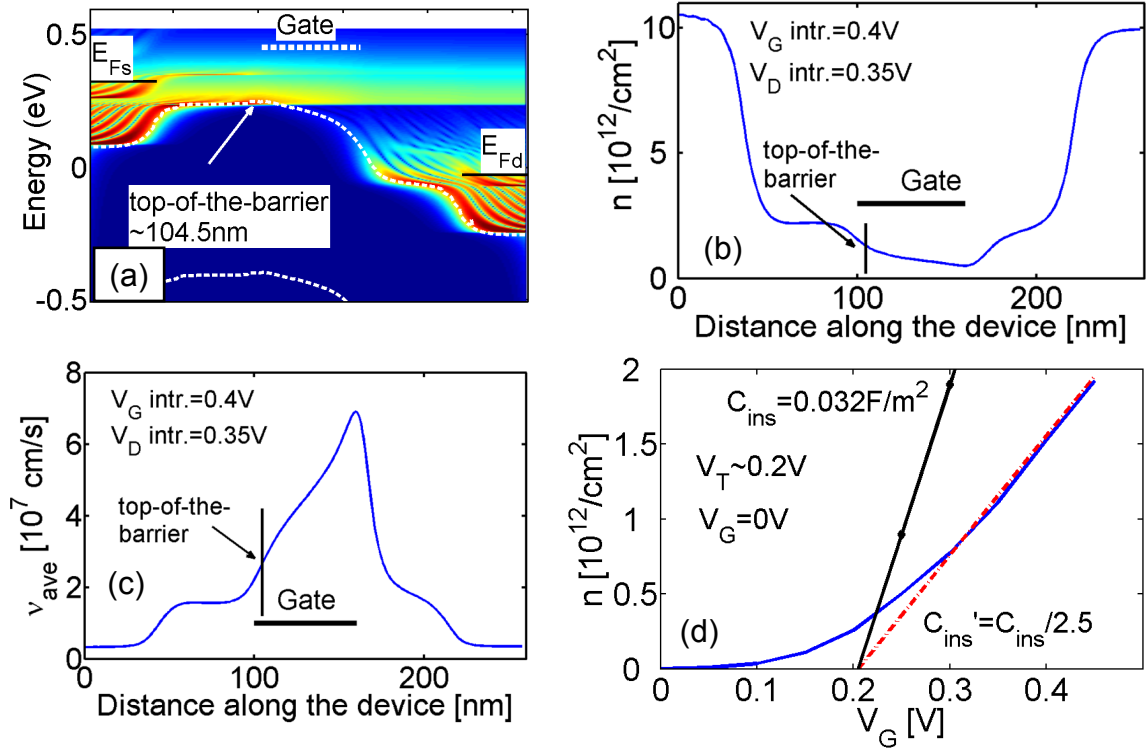

Figure 9

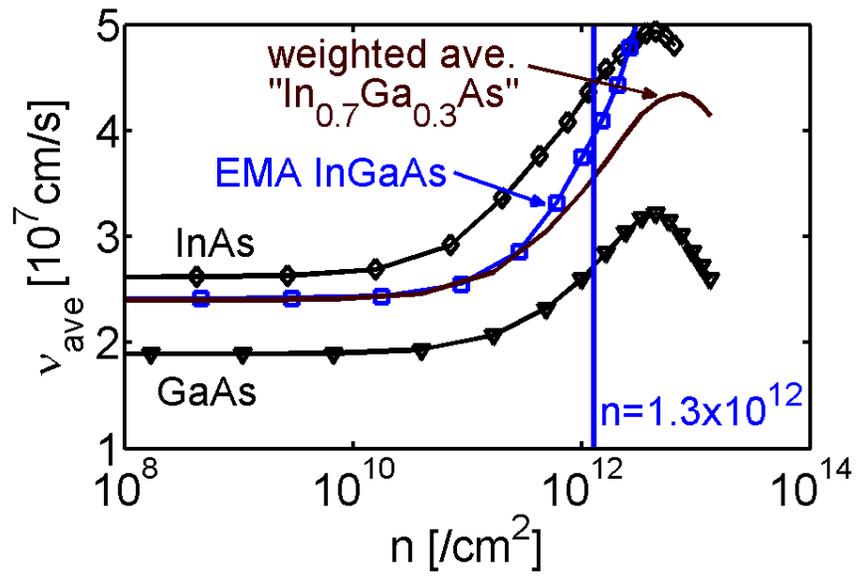

Figure 10: tins dependence

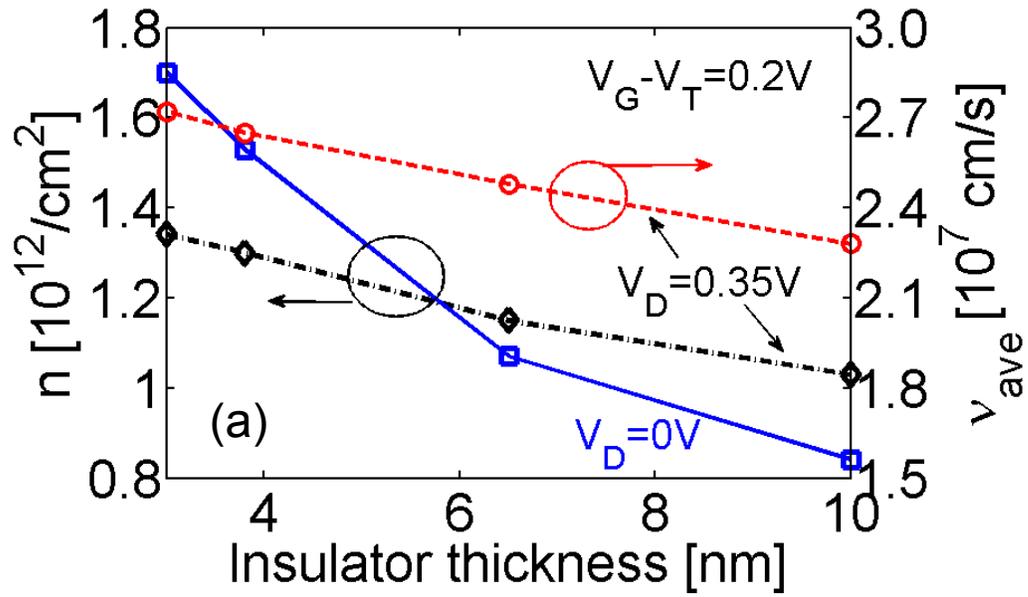

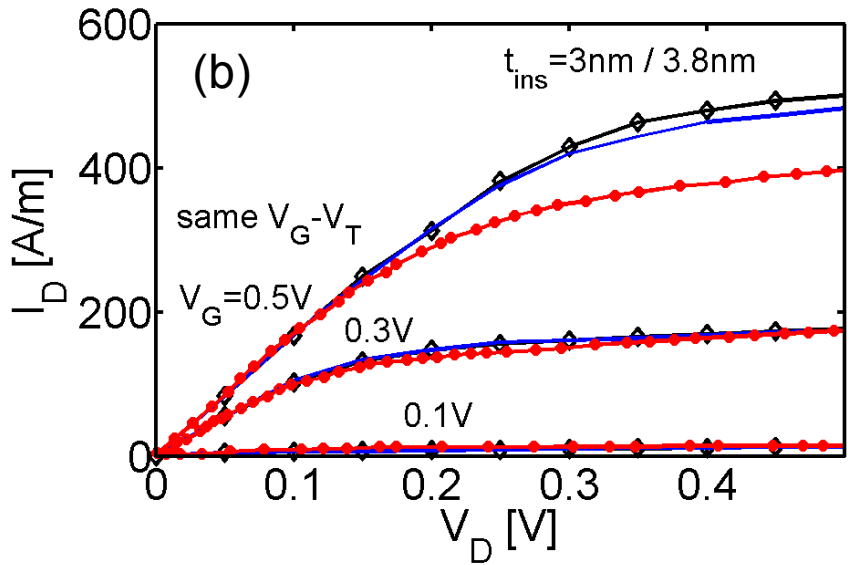

Figure 11: $L_G$ dependence

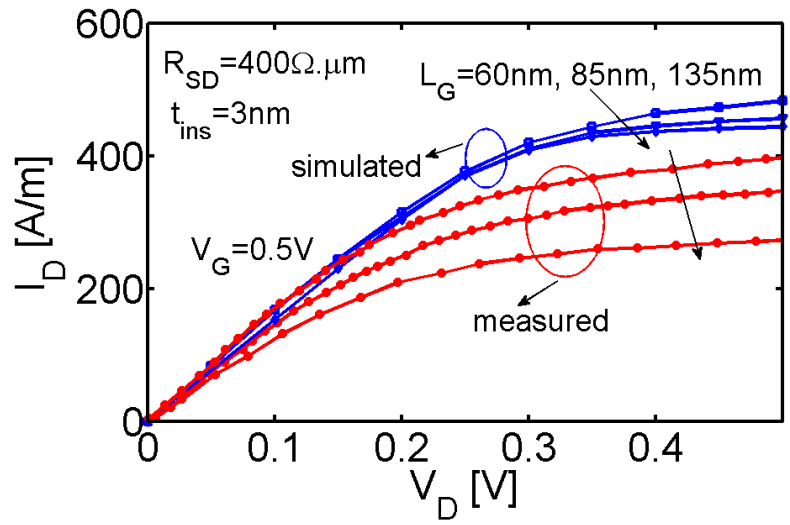